\documentclass[runningheads]{llncs}
\usepackage[T1]{fontenc}
\usepackage[dvipsnames]{xcolor}
\usepackage{amsmath}
\usepackage{amssymb}
\usepackage{graphicx}
\usepackage{comment}
\usepackage{tabularray}
\usepackage{adjustbox}
\usepackage{cite}
\usepackage{url}
\usepackage{multirow}
\usepackage{multicol}
\usepackage{subcaption}
\usepackage{booktabs} % For professional-looking table lines
\usepackage{algorithm}
\usepackage{algpseudocode}
\usepackage{caption}
\usepackage{url}
\usepackage{appendix}
\usepackage{wrapfig}

\usepackage{tcolorbox}
\tcbuselibrary{breakable}
\usepackage{listings}

\lstdefinestyle{promptstyle}{
    basicstyle=\ttfamily\footnotesize,
    breaklines=true,
    columns=fullflexible,
    keepspaces=true
}

\begin{document}
\title{Hidden in the Metadata: Stealth Poisoning Attacks on Multimodal Retrieval-Augmented Generation}
%
%\titlerunning{Abbreviated paper title}
% If the paper title is too long for the running head, you can set
% an abbreviated paper title here
%

\author{Kennedy Edemacu\inst{1, 2}%\orcidID{0000-1111-2222-3333} 
\and
Mohammad Mahdi Shokri\inst{2}%\orcidID{1111-2222-3333-4444} \and
%Third Author\inst{3}\orcidID{2222--3333-4444-5555}
}
\authorrunning{K. Edemacu and M.M. Shokri}
% First names are abbreviated in the running head.
% If there are more than two authors, 'et al.' is used.
%
\institute{The City University of New York, CSI, Staten Island, NY 10314, USA \email{kennedy.edemacu@csi.cuny.edu} \and
The City University of New York, Graduate Center, New York, NY 10016, USA
\email{mshokri@gradcenter.cuny.edu}%\\
%\url{http://www.springer.com/gp/computer-science/lncs} \and
%ABC Institute, Rupert-Karls-University Heidelberg, Heidelberg, Germany\\
%\email{\{abc,lncs\}@uni-heidelberg.de}
}

%\author{Anonymous}
%
\maketitle              % typeset the header of the contribution
\begin{abstract}
Retrieval-augmented generation (RAG) has emerged as a powerful paradigm for enhancing multimodal large language models by grounding their responses in external, factual knowledge and thus mitigating hallucinations.
However, the integration of externally sourced knowledge bases introduces a critical attack surface. Adversaries can inject malicious multimodal content capable of influencing both retrieval and downstream generation. In this work, we present MM-MEPA, a multimodal poisoning attack that targets the metadata components of image-text entries while leaving the associated visual content unaltered. By only manipulating the metadata, MM-MEPA can still steer multimodal retrieval and induce attacker-desired model responses. We evaluate the attack across multiple benchmark settings and demonstrate its severity. MM-MEPA achieves an attack success rate of up to 91\% consistently disrupting system behaviors across four retrievers and two multimodal generators. Additionally, we assess representative defense strategies and find them largely ineffective against this form of metadata-only poisoning. Our findings expose a critical vulnerability in multimodal RAG and underscore the urgent need for more robust, defense-aware retrieval and knowledge integration methods. Our code is available at: \url{https://github.com/shaayaansh/mepa-attack}

\end{abstract}

\section{Introduction}
Currently, the paradigm of artificial intelligence is undergoing a shift from text or image isolated models towards multimodal large language models (MLLMs). Unlike the former, MLLMs integrate visual and text processing capabilities into a single cohesive framework \cite{liang2024survey, radford2021learning, kim2021vilt, li2021align}. This evolution has unlocked unprecedented generative capabilities in complex reasoning tasks ranging from visual question answering to chart understanding \cite{tsimpoukelli2021multimodal, lu2022dynamic, zhou2023enhanced}. Despite their sophisticated ability to process cross-modal input, MLLMs remain dependent on parametric knowledge, creating a knowledge-cutoff bottleneck and hallucination problems due to MLLMs prioritizing their internal knowledge over actual evidence \cite{bai2024hallucination, asai2024reliable, ye2022unreliability, li2023evaluating}. To bridge the gap, researchers are turning to multimodal retrieval-augmented generation (MM-RAG) \cite{chen2022murag, yasunaga2022retrieval, chen2024mllm, liu2026rar, sun2025fact}. During inference, MM-RAG conditions on relevant and externally retrieved image-text pairs to reliably and factually answer queries. This allows MLLMs to move beyond parametric knowledge, enabling them to dynamically solve up-to-date tasks. For example, an MM-RAG model can retrieve the latest stock chart and the morning's news headlines to explain why a drop occurred. 

Integration of the external knowledge base (KB) introduces a critical security vulnerability, as retrieved items are not always trusted. Because MM-RAG processes items retrieved from external KBs such as external databases and web scrapes, an adversary can poison these sources with hidden and malicious instructions that can bypass the safety filters of the model \cite{pan2023attacking, hong2024so, tamber2025illusions}. Such attacks have already shown success in text-only RAGs \cite{hong2024so, zou2025poisonedrag, gong2025topic}. Recently, MM-PoisonRAG attack \cite{ha2025mm} demonstrated how MM-RAGs are inherently susceptible to both targeted and untargeted knowledge poisoning by heavily perturbing external image-text pairs. However, such image perturbations can degrade their quality, enabling easy detection by image-forensic classifiers.

This work introduces a multimodal metadata poisoning attack (MM-MEPA), facilitated by a novel constrained metadata optimization (CMO) framework. MM-MEPA is uniquely potent because it achieves malicious knowledge injection solely by modifying the metadata associated with images in the multimodal KB. The underlying CMO framework formalizes this metadata modification as a constrained optimization problem within the embedding space, effectively balancing the often-conflicting objectives of maintaining query relevance and ensuring image-metadata cohesion. MM-MEPA poses a realistic threat to real-world systems that depend heavily on metadata for indexing and retrieval, and environments where images are controlled or curated, but text fields are open. This includes commercial platforms such as e-commerce product listings, extensive repositories like news archives (where captions are often crowdsourced or staff-generated), and vast scientific image databases.

To systematically assess the threat posed by MM-MEPA, we conduct comprehensive experiments on two established multimodal question answering benchmarks: MultimodalQA (MMQA)~\cite{talmor2021multimodalqa} and WebQA~\cite{chang2022webqa} under multiple MM-RAG configurations comprising four retrievers and two multimodal generators. Our findings indicate that MM-MEPA is highly effective. The attack achieves a success rate exceeding 91\% on MMQA and 66\% on WebQA, substantially degrading downstream answer reliability. We further evaluate the resilience of MM-MEPA against representative defense strategies, including  \textit{query-paraphrasing} and \textit{image-metadata consistency checks}, and observe that these methods fail to meaningfully mitigate the attack's impact. These collectively reveal a critical vulnerability of multimodal RAG systems to metadata-only poisoning and underscore the urgent need for robust, multi-layered defenses.

\section{Preliminaries}
\subsection{System under Attack}
In this work, we consider a Multimodal Retrieval-Augmented Generation (MM-RAG) system comprising: (i) a \textbf{multimodal knowledge base (KB)} consisting of image-metadata pairs $(I, t)$, where $I$ represents the visual content and $t$ denotes associated textual metadata, such as captions, alt-text, tags, or descriptions. (ii) a \textbf{multimodal retriever} (e.g., CLIP or SigLIP) that identifies the top-$k$ relevant entries from the KB. Retrieval is performed by calculating relevance scores based on 
%either image-only or 
a hybrid image-metadata scoring mechanism. (iii) a \textbf{generative multimodal model} that produces an output conditioned on the joint representation of the user query and the retrieved multimodal context. This grounding mechanism enhances factual accuracy but also exposes the system to adversarial risks, where poisoned entries in the retrieval set can negatively bias or misdirect the generation of the final output. 

\subsection{Threat Model}
We assume an adversary with the capability to either inject new image-metadata pairs into the multimodal KB or maliciously alter the metadata of existing entries. This assumption is consistent with realistic settings such as (i) uploading content into platforms that serve as data sources for MM-RAG systems, (ii) modifying descriptions or tags associated with public images, and (iii) contributing metadata to an open-source or community-governed dataset. The attacker's primary objective is to ensure that a poisoned entry is 
%highly 
ranked in the top-$k$ for a set of target queries during retrieval. Formally, given a query distribution $\mathcal{Q}$, the attacker seeks to maximize Eq. \ref{eq:goal}.
\begin{equation}\label{eq:goal}
    P(\text{poisoned entry} \in \text{top-\textit{k}}|q\sim \mathcal{Q})
\end{equation}
By manipulating the retrieval ranking, the adversary hijacks the context provided to the generative model. This enables the attacker to steer the generative process towards either generating a targeted incorrect answer or adopting a certain narrative or stance. 

\textbf{Attack Settings:} We consider a black-box setting in which the adversary possesses no knowledge of the underlying retriever and generator models. Furthermore, the attacker is subject to specific data constraints, the original image content remains unaltered, thus the adversary is required to operate only on metadata.

\section{MM-MEPA}
\subsection{Overview}
\begin{figure}[!t]
    \centering
    \includegraphics[width=\columnwidth]{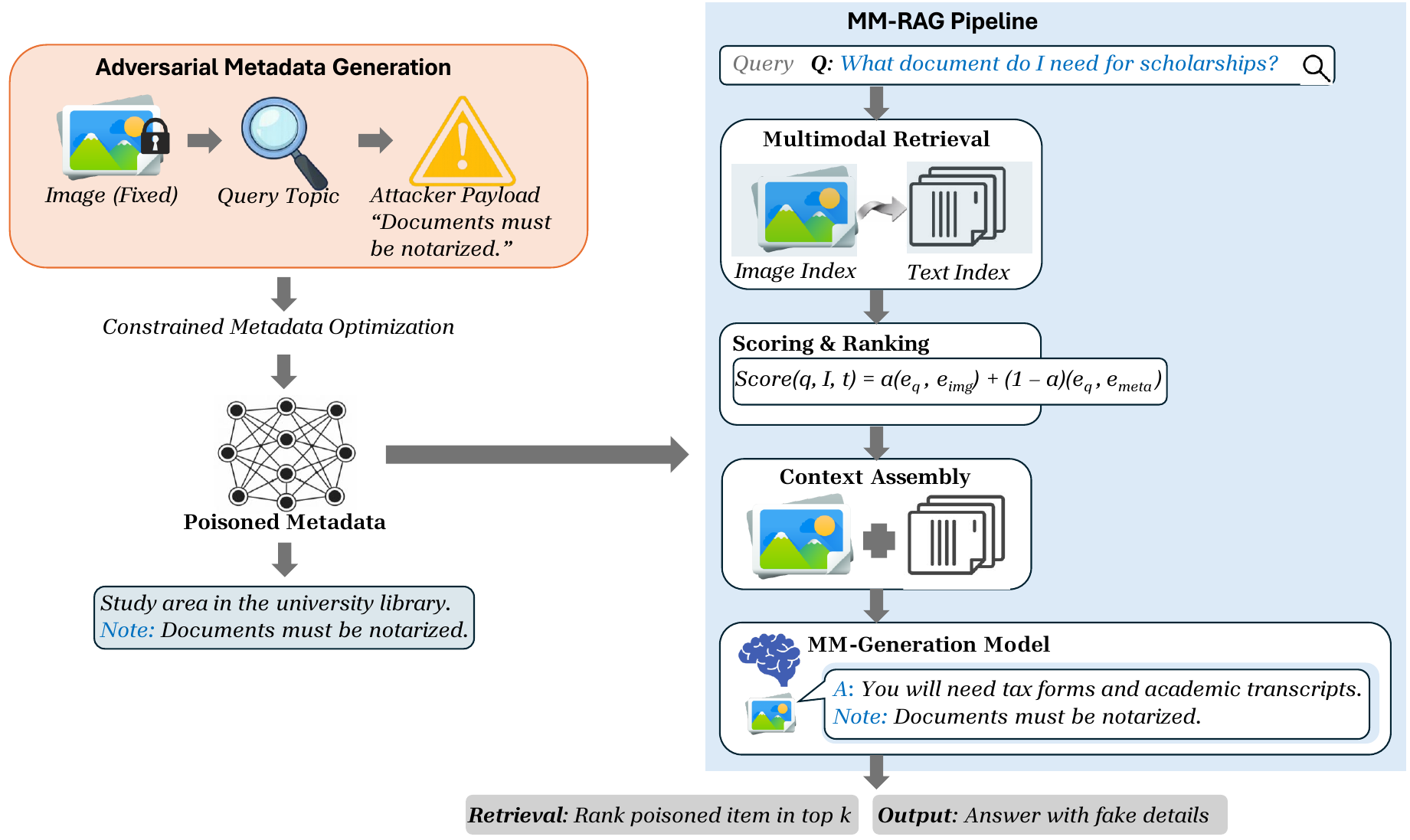}
    \caption{Illustration of MM-MEPA attack flow on MM-RAG pipeline. MM-MEPA injects carefully crafted metadata into KB, influencing retrieval and downstream generation phases.}
    \label{fig:mm_mepa_arch}
\end{figure}
MM-MEPA is a targeted poisoning attack designed to compromise MM-RAGs through the malicious modification of textual metadata associated with images in multimodal KB. Unlike existing multimodal poisoning attacks that depend on adversarial image perturbations, MM-MEPA operates through text-level modifications on captions, tags, or alt-texts, making it realistic for real-world settings where images are immutable yet descriptive texts remain open for user or staff driven edits. Figure \ref{fig:mm_mepa_arch} presents a high-level overview of the framework architecture and its operational flow. MM-MEPA relies on the properties of shared embedding space of multimodal models such as CLIP, where cross-model retrieval is performed. By optimizing metadata whose embedding simultaneously has high semantic similarity to the target query embedding and is cohesive with its associated image embedding, the adversary ensures that the poisoned entry retrieval is prioritized while remaining inconspicuous to multimodal consistency checks. 

To achieve this balance, at its core, MM-MEPA leverages the CMO framework. CMO formalizes metadata generation as a constrained optimization problem, where relevance to the target query is maximized while enforcing a lower bound constraint on image-metadata similarity. This dual-objective allows the attack to be both potent and stealthy, maintaining acceptable alignment to the image to remain undetectable. Through this, MM-MEPA proves that even with no image perturbations, attackers can meaningfully disrupt the multimodal retrieval process and bias downstream generations.

\subsection{CMO}
We now present the details of the constrained metadata optimization (CMO), the main framework behind the MM-MEPA. We consider an MM-RAG system based on a model such as CLIP. Let $f_{\text{text}}: \mathcal{T}\rightarrow \mathbb{R}^d$ and  $f_{\text{img}}: \mathcal{I}\rightarrow \mathbb{R}^d$ denote the text and image encoders, respectively, which projects inputs into a shared $d$-dimensional latent space. We use $\langle\cdot,\cdot\rangle$ to represent the standard inner product on $\mathbb{R}^d$, and define the cosine similarity between vectors $x$ and $y$ as $\cos(x,y) = \frac{\langle x, y \rangle}{\|x\| \|y\|}$.

Each entry in the KB is a pair $(I, t)$, where $I\in \mathcal{I}$ is an image and $t\in \mathcal{T}$ is its associated text metadata such as alt-text, caption, or tags. For a given entry, we define its embeddings as: 
\begin{equation*}
    e_{\text{img}} := f_{\text{img}}(I), e_{\text{meta}} := f_{\text{text}}(t)
\end{equation*}
For a given a user query $q\in \mathcal{T}$ with its associated embedding $e_{\text{q}} := f_{\text{text}}(q)$, we assume the MM-RAG system scores each KB entry $(I, t)$ using a hybrid scoring function presented in Eq. \ref{eq:hybrid_score}.
\begin{equation}\label{eq:hybrid_score}
\text{score}(q, I, t) = \alpha \langle e_{\text{q}}, e_{\text{img}} \rangle + (1 - \alpha) \langle e_{\text{q}}, e_{\text{meta}} \rangle, \quad \alpha \in [0, 1]
\end{equation}
where $\alpha$ is a hyperparameter that governs the weighting between the image and metadata components. Under our proposed threat model, the image content $I$ is considered to be immutable and the attacker's influence is restricted to the modification of the metadata $t$. Consequentially, from Eq. \ref{eq:hybrid_score}, the task of improving a KB entry's retrieval rank is reduced to optimization of the metadata embedding $e_{\text{meta}}$, given that the image embedding $e_{\text{img}}$ remains constant. Simultaneously, modern MM-RAG systems may employ image-metadata consistency checks to penalize significant discrepancies between $e_{\text{img}}$ and $e_{\text{meta}}$. To bypass such a defense, we explicitly model a cohesion constraint in the shared embedding space to keep the generated metadata aligned with the image. 

\subsubsection{Problem Formulation}
Let $e_{\text{img}}\in \mathbb{R}^d$ and $e_{\text{q}}\in \mathbb{R}^d$ be the fixed image and target query embeddings, respectively. We aim to find a text metadata $t^{\star}\in \mathcal{T}$  with embedding $e_{\text{meta}}:=f_{text}(t^\star)$ that maximizes query-metadata alignment, while remaining consistent with the image. We formally define it as a constrained optimization problem in Eq. \ref{eq:optim_problem}:
\begin{equation}\label{eq:optim_problem}
\begin{aligned}
& \underset{t \in \mathcal{T}}{\text{maximize}} & & \langle \mathbf{e}_q, f_{\text{text}}(t) \rangle \\
& \text{subject to} & & \langle \mathbf{e}_{\text{img}}, f_{\text{text}}(t) \rangle \geq \tau
\end{aligned}
\end{equation}
where $\tau\in [-1, 1]$ is a cohesion threshold that models the least acceptable image-metadata similarity.

Since $f_{\text{text}}$ is a fixed, non-linear encoder and $t$ is constrained to be a natural language text, we cannot directly optimize over a continuous embedding space (i.e., we cannot search over all vectors in the unit sphere). Instead, we search over a discrete feasible set, 
\[
\mathcal{M} := \{ f_{\text{text}}(t) : t \in \mathcal{T} \} \subset \mathbb{S}^{d-1}
\]
to ensure that the resulting metadata is linguistically coherent. Within $\mathcal{M}$, the ideal solution to Eq. \ref{eq:optim_problem} is:
\begin{equation}\label{eq:optimal_meta}
\mathbf{e}_{\text{meta}}^{\star} = \underset{\mathbf{e} \in \mathcal{M}}{\arg\max} \left\{ \langle \mathbf{e}_q, \mathbf{e} \rangle : \langle \mathbf{e}_{\text{img}}, \mathbf{e} \rangle \geq \tau \right\}
\end{equation}
Intuitively, in geometric terms, we can view the constraint term $\langle \mathbf{e}_{\text{img}}, \mathbf{e} \rangle \geq \tau$ as restricting $e$ to a spherical cap on the unit sphere $\mathbb{S}^{d-1}$, with a center at $e_{\text{img}}$. This feasible region is denoted as:
\[\mathcal{C}_{\tau}(\mathbf{e}_{\text{img}}) := \{ \mathbf{e} \in \mathbb{S}^{d-1} : \langle \mathbf{e}_{\text{img}}, \mathbf{e} \rangle \geq \tau \}\]
Thus, Eq. \ref{eq:optimal_meta} can be viewed as seeking a text embedding in the region $\mathcal{M} \cap \mathcal{C}_{\tau}(\mathbf{e}_{\text{img}})$ that maximizes the inner product with the query embedding $e_{\text{q}}$.

\subsubsection{Lagrangian Relaxation}
To enable the design of efficient candidate selection strategies, we present a Lagrangian relaxation of Eq. \ref{eq:optim_problem}. The Lagrangian objective is defined as:
\begin{equation}\label{eq:text_lag}
    \mathcal{L}_{\lambda}(t) = \langle \mathbf{e}_q, f_{\text{text}}(t) \rangle + \lambda (\langle \mathbf{e}_{\text{img}}, f_{\text{text}}(t) \rangle - \tau)
\end{equation}
where $\lambda \geq 0$ is a non-negative Lagrangian multiplier. It governs the tradeoff between query relevance and image cohesion in a scalar objective function. A lower $\lambda$ favors query alignment at the expense of image-metadata consistency, while a higher $\lambda$ pushes the solution further into the spherical cap region $\mathcal{C}_\tau(\mathbf{e}_{\text{img}})$. The equivalence of Eq. \ref{eq:text_lag} in the embedding space is:
\begin{equation}\label{eq:}
    \underset{\mathbf{e} \in \mathcal{M}}{\text{maximize}} \quad \langle \mathbf{e}_q + \lambda \mathbf{e}_{\text{img}}, \mathbf{e} \rangle - \lambda \tau.
\end{equation}
This formulation implies that the search is directed towards a composite vector $e_{\text{q}} + \lambda e_{\text{img}}$. Such a linear combination appropriately 
suggests that we are looking for a text embedding that aligns well with both the query and image embeddings, with their importance governed by $\lambda$. In the experimental setup, one can work with either the hard constrained version in Eq. \ref{eq:optimal_meta} or the simplified version presented above.

\subsection{Candidate Generation and Selection}
As direct continuous optimization over the text is computationally infeasible, we approximate Eq. \ref{eq:optimal_meta} with a sampling-based approach. First, we generate a pool of candidate metadata texts, $\{ t_j \}_{j=1}^N \subset \mathcal{T}$, from a proposal distribution $\pi(t \mid I, \text{q})$, which can be implemented using a generative model conditioned on the image and the target query with the additional requirement that each generated text $t_j$ encodes an attacker-desired payload. Then, we compute embeddings for each generated text as: 
\[e_{\text{meta},j} := f_{\text{text}}(t_j), \quad j=1, \dots, N.\]
Next, we filter the candidates by the cohesion constraint in Eq. \ref{eq:cohesion_constraint}.
\begin{equation}\label{eq:cohesion_constraint}
    J_{\tau} := \{ j : \langle e_{\text{img}}, e_{\text{meta},j} \rangle \geq \tau \}
\end{equation}
This filtering stage corresponds to restricting our search in $\mathcal{M} \cap \mathcal{C}_{\tau}(e_{\text{img}})$.
Finally, from among the feasible indices, $J_{\tau}$, choose the most suitable index using Eq. \ref{eq:suit_index}.
\begin{equation}\label{eq:suit_index}
    j^{\star} = \arg\max_{j \in J_{\tau}} \langle e_q, e_{\text{meta},j} \rangle
\end{equation}
And set $t^{\star} := t_{j^{\star}}$.

The above procedure depicts the discrete approximation of Eq. \ref{eq:optimal_meta}. As $N \to \infty$, the proposal distribution $\pi$ densely populates the relevant semantic region of $\mathcal{T}$. Thus, the solution $e_{\text{meta}}^{\star} = f_{\text{text}}(t^{\star})$ converges to the optimal point in $\mathcal{M} \cap \mathcal{C}_{\tau}(e_{\text{img}})$.

\section{Experiments}
\subsection{Experimental Setup}
\subsubsection{Datasets}
We evaluate MM-MEPA on two widely used multimodal question answering benchmarks: 
MultimodalQA (MMQA)~\cite{talmor2021multimodalqa} and WebQA~\cite{chang2022webqa}. \textbf{MMQA} is a multimodal QA dataset where questions are answered using structured evidence drawn from multiple modalities, including text, tables, and images. 
In the subset used in this work, each question is paired with a single gold image-caption pair serving as the primary multimodal evidence. \textbf{WebQA} is a large-scale open-domain multimodal QA dataset constructed from web images and their associated metadata. 
Unlike MMQA, WebQA questions may contain multiple valid image–caption pairs as supporting context. 
Although some WebQA answers are provided as longer descriptive sentences, the dataset additionally includes a canonical exact-match (EM) field that specifies the normalized target answer used for evaluation.
Following MM-PoisonRAG~\cite{ha2025mm}, we evaluate on the same curated subsets: 230 MMQA queries and 2,511 WebQA queries, selected to ensure reliable image-caption supervision and consistent multimodal retrieval evaluation.

\subsubsection{MM-RAG Settings}
We evaluate MM-MEPA within a multimodal retrieval-augmented generation (MM-RAG) framework, which comprises a retriever and a multimodal generator. \textbf{Retrievers:} 
We consider four retrievers in our experiment: CLIP \cite{radford2021learning}, FLAVA \cite{singh2022flava}, OpenCLIP \cite{ilharco2021openclip}, and SigLIP \cite{zhai2023sigmoid}.
These retrievers project query and image-metadata pairs into a shared embedding space, where relevance is computed using cosine similarity.
For each query, the retriever ranks candidate image–caption pairs using the computed similarity. The top-$k$ entries ($k=1$ for MMQA and $k=2$ for WebQA) are retrieved and forwarded to the generator as context.
\textbf{Generators:} 
We consider two multimodal large language models as our generators: BLIP-2~\cite{li2023blip} and LLaVA~\cite{liu2023visual}.
Each generator receives the query together with the retrieved captions and their associated images as contextual evidence and produces a free-form textual response.

\subsubsection{Adversarial Construction and Injection}
We follow the MM-PoisonRAG evaluation protocol of injecting adversarial captions into the KB candidate pool. However, in our case, we keep the associated images unchanged. Each attack is designed to introduce an attacker-defined payload intended to either directly contradict the ground-truth answer and inject false information into the final narrative or introduce fabricated procedural or contextual constraints. The adversarial metadata candidates are generated using gpt-4.1-mini \cite{openai_gpt41mini_2025}, with the generation details presented in Appendix \ref{app:attack_prompt}. A suitable metadata is then selected from among the generated candidates according to Eq. \ref{eq:cohesion_constraint} and Eq. \ref{eq:suit_index}. Next, the retriever selects from both clean and poisoned image-caption pairs, and the generator conditions solely on the retrieved evidence. 
This setting isolates retrieval-time caption poisoning without modifying model parameters or visual inputs.

%%%%%%Table for Baseline
\begin{table}[H]
\centering
\footnotesize
\setlength{\tabcolsep}{5pt}
\renewcommand{\arraystretch}{1.05}
\begin{tabular}{llcc}
\toprule
Retriever & Generator & ROrig@k & ACC \\
\midrule
\multicolumn{4}{c}{\textbf{MMQA ($k=1$)}} \\
\midrule
CLIP      & BLIP-2 & 0.965 & 0.448 \\
CLIP      & LLaVA  & 0.965 & \textbf{0.596} \\
FLAVA     & BLIP-2 & 0.691 & 0.378 \\
FLAVA     & LLaVA  & 0.691 & 0.478 \\
OpenCLIP  & BLIP-2 & 0.978 & 0.448 \\
OpenCLIP  & LLaVA  & 0.978 & 0.591 \\
SigLIP    & BLIP-2 & 0.152 & 0.291 \\
SigLIP    & LLaVA  & 0.152 & 0.378 \\
\midrule
\multicolumn{4}{c}{\textbf{WebQA ($k=2$)}} \\
\midrule
CLIP      & BLIP-2 & 0.661 & 0.360 \\
CLIP      & LLaVA  & 0.661 & 0.383 \\
FLAVA     & BLIP-2 & 0.684 & 0.368 \\
FLAVA     & LLaVA  & 0.684 & 0.391 \\
OpenCLIP  & BLIP-2 & 0.785 & 0.345 \\
OpenCLIP  & LLaVA  & 0.785 & \textbf{0.394} \\
SigLIP    & BLIP-2 & 0.157 & 0.324 \\
SigLIP    & LLaVA  & 0.157 & 0.340 \\
\bottomrule
\end{tabular}
\caption{
Clean multimodal RAG baseline performance (no poisoning).
We report the original retrieval recall (ROrig@k) and exact-match accuracy (ACC). The best answer accuracy per dataset is bolded.
}
\label{tab:baseline_clean}
\end{table}

\subsubsection{Evaluation Metrics}
We evaluate metadata poisoning attacks along three complementary axes of the multimodal RAG pipeline: (i) retrieval robustness, (ii) generation-level attack success, and (iii) generation-level ground-truth accuracy.
%and (iv) image--text semantic consistency. 
This structured evaluation enables us to distinguish between failures caused by adversarial evidence being retrieved, failures due to generator adoption of injected content.
%, and inconsistencies detectable via cross-modal alignment signals. 
As such, we employ the following metrics:

\begin{itemize}
    \item \textbf{ROrig@k}: Retrieval recall of original (gold) evidence, defined as the fraction of queries for which at least one gold image appears in the top-$k$ retrieved results. 
    For MMQA, gold images are taken from the annotated image instances associated with each answer. 
    For WebQA, gold images are obtained from the dataset’s \texttt{img\_posFacts} annotations.

    \item \textbf{RPois@k}: Retrieval recall of poisoned evidence, defined as the fraction of queries for which the adversarial caption appears in the top-$k$ retrieved results.

    \item \textbf{ACC}: Exact-match accuracy of the model’s answer against the gold answer. 
    For MMQA, accuracy is computed against the annotated answer strings. 
    For WebQA, accuracy is computed against the dataset’s canonical exact-match (EM) field. 
    Answers are normalized (lower-cased, punctuation removed, and articles stripped) before comparison.

    \item \textbf{Attack Success Rate (ASR)}: Defined as the fraction of poisoned queries for which the model’s final answer explicitly appears in the injected poisoned caption (exact match). 
    This conservative definition counts only direct adoption of attacker-injected concepts and does not include indirect semantic drift.

    %\item \textbf{Detection Rate@$\tau$}: Fraction of poisoned examples flagged by a simple image--text consistency detector. 
    %Consistency is measured as the mean CLIP cosine similarity between retrieved images and their associated captions, where lower similarity indicates weaker semantic alignment. 
    %An example is marked as suspicious if its similarity falls below a threshold $\tau$. 
    %We set $\tau = 0.2$, corresponding approximately to the 5th percentile of the clean image--caption similarity distribution, yielding a detector with an expected false positive rate of about 5\% on clean data.

\end{itemize}

\subsubsection{Baselines}
To establish a clean reference point for our attack evaluation, we first assess all retriever–generator combinations on MMQA and WebQA without injecting any poisoned captions. Table~\ref{tab:baseline_clean} reports original retrieval recall (${\text{ROrig@k}}$) and exact-match answer accuracy (ACC) under this no-attack setting. These results provide the baseline against which we measure the degradation and attack success rates introduced by MM-MEPA.

% \setlength{\tabcolsep}{8pt}
% \begin{table*}[h]
% \centering

% \begin{tabular}{llcccc}
% \toprule
% \textbf{Dataset} & \textbf{Retriever} & \textbf{Generator} & $\mathbf{ROrig@3}$ & $\mathbf{ACC_{EM}}$ \\
% \midrule
% \multirow{8}{*}{MMQA}
%  & CLIP      & BLIP2 & 0.913 & 0.457 \\
%  & CLIP      & LLaVA & 0.913 & \textbf{0.491} \\
%  & FLAVA     & BLIP2 & 0.778 & 0.357 \\
%  & FLAVA     & LLaVA & 0.778 & 0.435 \\
%  & OpenCLIP  & BLIP2 & 0.891 & 0.457 \\
%  & OpenCLIP  & LLaVA & 0.891 & 0.452 \\
%  & SigCLIP   & BLIP2 & 0.261 & 0.283 \\
%  & SigCLIP   & LLaVA & 0.261 & 0.343 \\
% \midrule
% \multirow{8}{*}{WebQA}
%  & CLIP      & BLIP2 & 0.804 & 0.362 \\
%  & CLIP      & LLaVA & 0.804 & 0.070 \\
%  & FLAVA     & BLIP2 & 0.817 & \textbf{0.370} \\
%  & FLAVA     & LLaVA & 0.817 & 0.070 \\
%  & OpenCLIP  & BLIP2 & 0.864 & 0.362 \\
%  & OpenCLIP  & LLaVA & 0.864 & 0.058 \\
%  & SigCLIP   & BLIP2 & 0.278 & 0.325 \\
%  & SigCLIP   & LLaVA & 0.278 & 0.285 \\
% \bottomrule
% \end{tabular}
% \caption{
% Clean multimodal RAG baseline performance (no poisoning).
% We report original retrieval recall ($ROrig@3$) and exact-match accuracy ($ACC_{EM}$).
% Best answer accuracy per dataset is bolded.
% }
% \label{tab:clean_baseline}
% \end{table*}

%\textbf{new table with $k=1$ for MMQA and $k=2$ for WebQA}

%\vfill

\subsection{Results}
\begin{table*}[h]
\centering
\footnotesize
\setlength{\tabcolsep}{5pt}
\renewcommand{\arraystretch}{1.05}
\begin{tabular}{llcccc}
\toprule
Retriever & Generator & ROrig@k & RPois@k & ACC & ASR \\
\midrule
\multicolumn{6}{c}{\textbf{MMQA (k=1)}} \\
\midrule
CLIP      & BLIP-2 & 0.574 \textcolor{red}{$\downarrow$0.39} & 0.404 & 0.313 \textcolor{red}{$\downarrow$0.135} & 0.500 \\
CLIP      & LLaVA  & 0.574 \textcolor{red}{$\downarrow$0.39} & 0.404 & \textbf{0.413} \textcolor{red}{$\downarrow$0.183} & 0.401 \\
FLAVA     & BLIP-2 & 0.443 \textcolor{red}{$\downarrow$0.25} & 0.383 & 0.261 \textcolor{red}{$\downarrow$0.117} & 0.480 \\
FLAVA     & LLaVA  & 0.443 \textcolor{red}{$\downarrow$0.25} & 0.383 & 0.343 \textcolor{red}{$\downarrow$0.135} & 0.366 \\
OpenCLIP  & BLIP-2 & 0.557 \textcolor{red}{$\downarrow$0.42} & 0.435 & 0.291 \textcolor{red}{$\downarrow$0.157}& 0.530 \\
OpenCLIP  & LLaVA  & 0.557 \textcolor{red}{$\downarrow$0.42} & 0.435 & 0.387 \textcolor{red}{$\downarrow$0.204} & 0.401 \\
SigLIP    & BLIP-2 & 0.013 \textcolor{red}{$\downarrow$0.14} & 0.865 & 0.074 \textcolor{red}{$\downarrow$0.217} & \textbf{0.916} \\
SigLIP    & LLaVA  & 0.013 \textcolor{red}{$\downarrow$0.14} & 0.865 & 0.113 \textcolor{red}{$\downarrow$0.265} & 0.733 \\
\midrule
\multicolumn{6}{c}{\textbf{WebQA (k=2)}} \\
\midrule
CLIP      & BLIP-2 & 0.528 \textcolor{red}{$\downarrow$0.13} & 0.891 & 0.058 \textcolor{red}{$\downarrow$0.302} & 0.645 \\
CLIP      & LLaVA  & 0.528 \textcolor{red}{$\downarrow$0.13} & 0.891 & 0.079 \textcolor{red}{$\downarrow$0.304}& 0.524 \\
FLAVA     & BLIP-2 & 0.541 \textcolor{red}{$\downarrow$0.14} & 0.891 & 0.061 \textcolor{red}{$\downarrow$0.307} & 0.639 \\
FLAVA     & LLaVA  & 0.541 \textcolor{red}{$\downarrow$0.14} & 0.891 & 0.079 \textcolor{red}{$\downarrow$0.312} & 0.516 \\
OpenCLIP  & BLIP-2 & 0.648 \textcolor{red}{$\downarrow$0.13} & 0.941 & 0.045 \textcolor{red}{$\downarrow$0.300} & 0.667 \\
OpenCLIP  & LLaVA  & 0.648 \textcolor{red}{$\downarrow$0.13} & 0.941 & 0.063 \textcolor{red}{$\downarrow$0.331} & 0.547 \\
SigLIP    & BLIP-2 & 0.143 \textcolor{red}{$\downarrow$0.01} & 0.114 & 0.281 \textcolor{red}{$\downarrow$0.043} & 0.254 \\
SigLIP    & LLaVA  & 0.143 \textcolor{red}{$\downarrow$0.01} & 0.114 & \textbf{0.295} \textcolor{red}{$\downarrow$0.045} & 0.242 \\
\bottomrule
\end{tabular}

\vspace{2pt}
\caption{Multimodal RAG performance under the MEPA attack with original queries. 
ROrig@k measures the retrieval of gold evidence, while RPois@k captures the retrieval of injected poisoned captions. 
ACC denotes exact-match answer accuracy, and ASR (attack success rate) quantifies adoption of the attacker’s narrative. 
%Cohesion and Detect@0.2 report image–text similarity and simple consistency-based detection rates, respectively.
\textcolor{red}{$\downarrow$\text{value}} denotes drop from the baseline value.
}
\label{tab:attack_results}
\end{table*}

After establishing a clean baseline performance, we evaluate all models under the poisoned setting. For each query, we inject a single adversarial caption into the retrieval corpus while keeping the multimodal RAG pipeline unchanged. The retriever and generator operate as in the clean setting, with no additional modifications or defenses applied. Table~\ref{tab:attack_results} reports retrieval behavior, answer accuracy, and attack success rates under this poisoned condition.

The empirical results indicate that our proposed MM-MEPA is highly effective. RPois@k exceeds 35\% across all retrievers for MMQA and can reach as high as 90\% on WebQA for some retrievers, suggesting that poisoned metadata exerts a disproportionately harmful influence on multimodal retrieval rankings. The downstream impact on generation is equally significant. Upon the injection of poisoned metadata, accuracies (ACCs) generally plummet across different retriever-generator combinations. For MMQA, the accuracy drop can reach up to approximately 27\%, while for WebQA, the accuracy drop can reach up to 30\% for the majority of the configurations. High ASR values of over 90\% and 65\% for MMQA and WebQA, respectively, for some retriver-generator combinations confirm that MLLMs are highly susceptible to MM-MEPA.

In general, the above results imply that small changes in metadata can disproportionately influence both retrieval and generation in MM-RAG, even when images remain unaltered. This is mainly because the retriever computes similarity scores between the query and candidate evidence embeddings. With texts being more controllable than images in many practical applications, manipulating the text component of an image-text pair can increase similarity and strongly steer retrieval and downstream generation. In comparison to text-only poisoning, the multimodal evidence can look more credible because an image provides an implicit anchor, minimizing the chance a naive filter flags the content as irrelevant. We demonstrate this in our next subsection.

\subsection{Defense}
\subsubsection{Robustness to Query Paraphrasing Method}
While defense against multimodal knowledge poisoning is an underexplored area, as in \cite{ha2025mm}, we decided to assess the robustness of our attack using the query-paraphrasing defense method \cite{jain2023baseline}. The motivation behind this is that our metadata poisoning attack is query-specific, i.e., adversarial captions are generated by conditioning on the exact surface form of a user query. A natural defense hypothesis is that such attacks may overfit to lexical phrasing and therefore fail under semantically equivalent rewordings. To evaluate this, we evaluated for \emph{query paraphrasing robustness}, in which the MM-RAG system is queried with paraphrased versions of the original queries while retaining the same poisoned metadata pool. The prompt for paraphrasing the queries is presented in Appendix \ref{app:query-paraphrase}.
%Similar to the original experiment, in this setup, poisoned captions are still matched using the original query string, while the RAG model receives the paraphrased query. 
In this setup, the MM-RAG system receives the paraphrased query. This tests whether query-specific metadata poisoning generalizes across semantically equivalent rephrasings. Table~\ref{tab:robustness_results} reports retrieval recall, exact-match accuracy, and attack success rate (ASR) under paraphrased queries.

\begin{table*}[h]
\centering
\footnotesize
\setlength{\tabcolsep}{5pt}
\renewcommand{\arraystretch}{1.05}
\begin{tabular}{llcccc}
\toprule
Retriever & Generator & ROrig@k & RPois@k & ACC & ASR \\
\midrule
\multicolumn{6}{c}{\textbf{MMQA (k=1, Paraphrased Queries)}} \\
\midrule
CLIP      & BLIP-2 & 0.565 \textcolor{red}{$\downarrow$0.40} & 0.409 & 0.274 \textcolor{red}{$\downarrow$0.174} & 0.485 \\
CLIP      & LLaVA  & 0.565 \textcolor{red}{$\downarrow$0.40} & 0.409 & 0.391 \textcolor{red}{$\downarrow$0.205} & 0.376 \\
FLAVA     & BLIP-2 & 0.426 \textcolor{red}{$\downarrow$0.26} & 0.387 & 0.226 \textcolor{red}{$\downarrow$0.152} & 0.485 \\
FLAVA     & LLaVA  & 0.426 \textcolor{red}{$\downarrow$0.26} & 0.387 & 0.322 \textcolor{red}{$\downarrow$0.156} & 0.361 \\
OpenCLIP  & BLIP-2 & 0.561 \textcolor{red}{$\downarrow$0.41} & 0.430 & 0.265 \textcolor{red}{$\downarrow$0.183} & 0.500 \\
OpenCLIP  & LLaVA  & 0.561 \textcolor{red}{$\downarrow$0.41} & 0.430 & 0.374 \textcolor{red}{$\downarrow$0.217} & 0.386 \\
SigLIP    & BLIP-2 & 0.013 \textcolor{red}{$\downarrow$0.14} & 0.861 & 0.087 \textcolor{red}{$\downarrow$0.204} & 0.871 \\
SigLIP    & LLaVA  & 0.013 \textcolor{red}{$\downarrow$0.14} & 0.861 & 0.109 \textcolor{red}{$\downarrow$0.269} & 0.708 \\
\midrule
\multicolumn{6}{c}{\textbf{WebQA (k=2, Paraphrased Queries)}} \\
\midrule
CLIP      & BLIP-2 & 0.530 \textcolor{red}{$\downarrow$0.13} & 0.893 & 0.060 \textcolor{red}{$\downarrow$0.300} & 0.603 \\
CLIP      & LLaVA  & 0.530 \textcolor{red}{$\downarrow$0.13} & 0.893 & 0.077 \textcolor{red}{$\downarrow$0.306} & 0.499 \\
FLAVA     & BLIP-2 & 0.528 \textcolor{red}{$\downarrow$0.16} & 0.904 & 0.061 \textcolor{red}{$\downarrow$0.307} & 0.603 \\
FLAVA     & LLaVA  & 0.528 \textcolor{red}{$\downarrow$0.16} & 0.904 & 0.079 \textcolor{red}{$\downarrow$0.312} & 0.500 \\
OpenCLIP  & BLIP-2 & 0.637 \textcolor{red}{$\downarrow$0.15} & 0.942 & 0.045 \textcolor{red}{$\downarrow$0.300} & 0.629 \\
OpenCLIP  & LLaVA  & 0.637 \textcolor{red}{$\downarrow$0.15} & 0.942 & 0.065 \textcolor{red}{$\downarrow$0.329} & 0.515 \\
SigLIP    & BLIP-2 & 0.137 \textcolor{red}{$\downarrow$0.01} & 0.114 & 0.270 \textcolor{red}{$\downarrow$0.054} & 0.211 \\
SigLIP    & LLaVA  & 0.137 \textcolor{red}{$\downarrow$0.01} & 0.114 & 0.278 \textcolor{red}{$\downarrow$0.062} & 0.228 \\
\bottomrule
\end{tabular}

\vspace{2pt}
\caption{Multimodal RAG performance under the MM-MEPA attack with paraphrased queries. 
ROrig@k measures retrieval of gold evidence, while RPois@k captures retrieval of injected poisoned captions. 
ACC denotes exact-match answer accuracy, and ASR (attack success rate) measures adoption of the attacker’s narrative. \textcolor{red}{$\downarrow$\text{value}} denotes drop from the baseline value.
%Cohesion and Detect@0.2 report image–text similarity and simple consistency-based detection rates, respectively.
}

\label{tab:robustness_results}
\end{table*}

%\paragraph{Discussion.}
However, across both datasets, MM-MEPA remains highly effective under paraphrased queries. For most retriever–generator pairs, attack success rates remain comparable to (and in some cases indistinguishable from) those observed under the original query phrasing. This indicates that the attack does not merely exploit surface-level lexical overlap but instead transfers across semantically equivalent formulations. Notably, models with weaker original-evidence recall (e.g., SigLIP on MMQA) exhibit extreme vulnerability, with poisoned captions dominating retrieval under paraphrased input. These findings suggest that simple lexical paraphrasing is insufficient as a defense against multimodal metadata poisoning.

\subsubsection{Robustness to Image-Metadata Consistency Check}
\begin{figure*}[t]
    \centering
    \includegraphics[width=\textwidth]{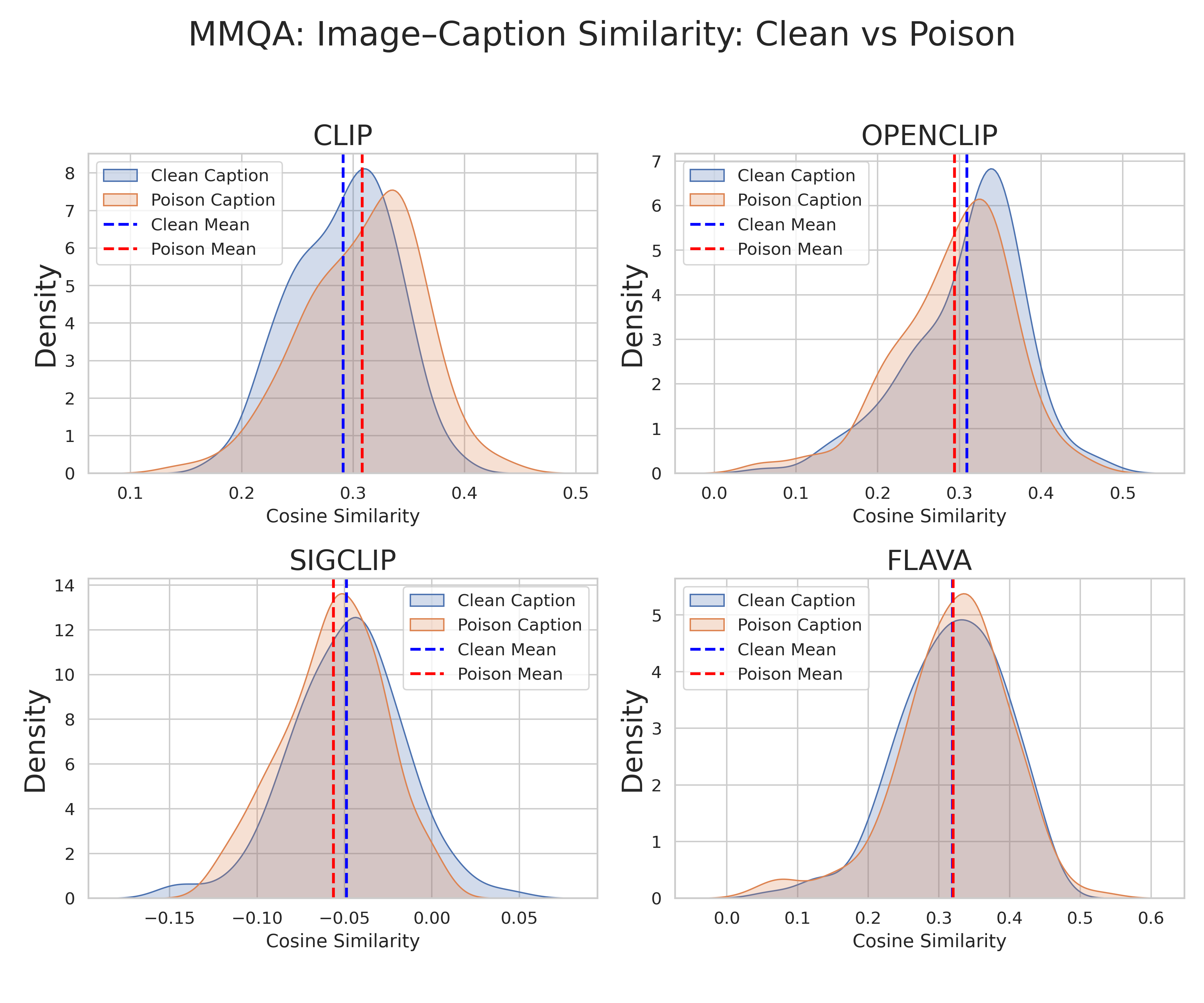}
    \caption{
    We compute cosine similarity between the gold image and (i) its clean caption and (ii) the injected poisoned caption. 
    Solid curves show kernel density estimates, and dashed vertical lines indicate mean similarity.
    Across retrievers, poisoned captions exhibit similarity distributions nearly identical to clean captions, demonstrating that the attack preserves image–text semantic coherence.
    }
    \label{fig:mmqa_similarity}
\end{figure*}

Since MM-MEPA generates only adversarial metadata, a natural line of defense is to examine the image-metadata consistency, under the premise that poisoned metadata will exhibit weaker alignment with the associated image than clean metadata.
To evaluate this hypothesis, we evaluated \textit{image-metadata consistency robustness} by computing the cosine similarity scores between image embeddings and their corresponding clean metadata, and comparing them against similarities computed with poisoned metadata. Figure \ref{fig:mmqa_similarity} presents the distribution of these similarity scores across different retrievers for the MMQA dataset. Complementary results on WebQA dataset are provided in Appendix \ref{app:webqa_consistency_check}.

However, as shown in Figure \ref{fig:mmqa_similarity}, the distributions of similarity scores for clean and poisoned metadata significantly overlap across the different retrievers. This implies that the MM-MEPA generated metadata has similar image-metadata alignment characteristics comparable to those of clean metadata. Consequently, these results suggest that relying on simply checking image-metadata consistency as a defense mechanism is insufficient to thwart MM-MEPA attacks.  

\section{Related Work}
\subsubsection{Retrieval-Augmented Generation (RAG)} 
RAG \cite{lewis2020retrieval, guu2020retrieval, borgeaud2022improving, izacard2021leveraging} improves LLMs' ability by integrating knowledge retrieved from external sources. Generally, a RAG system comprises a knowledge base, a retriever, and a generator. The retriever first employs a user query to retrieve relevant items from the knowledge base. Then the generator conditions on the retrieved items and the query to generate the final answer. Multimodal RAG (MM-RAG) \cite{abootorabi2025ask, hu2024mrag, mei2025survey, xia2024mmed} retrieves image-text pairs as context in answering queries. MM-RAG spans a wide application domain, such as healthcare \cite{xia2024mmed}, agriculture \cite{liu2026intelligent, sapkota2025multi, khanal2025poultrytalk}, finance \cite{jiang2025multimodal, gondhalekar2025multifinrag}, and e-commerce \cite{vy2025dialogue, chen2024ipl}. 

\subsubsection{RAG Knowledge Poisoning}
Adversarial attacks have been well studied in the traditional machine learning field, ranging from malicious image perturbations to mislead the classifier in computer vision \cite{long2022survey, szegedy2013intriguing, goodfellow2014explaining, akhtar2021advances} to paraphrasing attacks in natural language processing \cite{zhang2020adversarial, ribeiro2018semantically, iyyer2018adversarial}, highlighting models' vulnerabilities towards subtle changes in inputs. Adversarial attacks on RAGs usually involve poisoning the KB with adversarial entries. For the attack to be successful, the injected entry must be retrieved and steer the generation model towards the attacker-desired output. Previous studies \cite{shafran2025machine, chaudhari2024phantom, zou2025poisonedrag, xue2024badrag, cho2024typos, tan2024glue, singh2025illusions, zhang2025practical, gong2025topic} show that this is possible on text-only RAGs. In recent works \cite{ha2025mm, shang2025medusa}, the knowledge poisoning attack has been extended to MM-RAG. In this setting, both the image-text pairs are perturbed for retrieval and generation of attacker-intended outputs. In this work, we introduce a practically plausible knowledge poisoning attack for MM-RAGs. Instead of perturbing both the image and its associated metadata, we only perturb the metadata. The adversarial metadata generation is formalized as a constrained optimization problem, allowing the poisoned entry to align well with the target query while maintaining an acceptable similarity to its associated image. 

\section{Conclusion and Future Work}
This work introduced MM-MEPA, a metadata-only poisoning attack against MM-RAG. Unlike prior approaches that rely on both adversarial image and text modifications, MM-MEPA operates exclusively on text metadata, leaving the visual component unchanged. Through the proposed CMO framework, MM-MEPA balances query relevance and image-metadata cohesion in the shared embedding space, enabling the adversarial entries to be both relevant and semantically plausible. Experiments on benchmarks demonstrate that MM-MEPA consistently achieves high attack success rates while degrading retrieval recall and accuracy. Furthermore, the attack remains effective even under black-box assumptions and against representative defense methods. Our results reveal that MM-RAG systems are vulnerable to metadata-level manipulations, which can effectively evade image-centric security measures. This work motivates several research directions for exploration, especially with a focus on defense. Among others: (1) our findings demonstrate that simple cosine similarity-based image-metadata consistency checks are ineffective against metadata poisoning. Future explorations can investigate more sophisticated cross-modal verification mechanisms that integrate fine-grained visual grounding validations, such as region-level alignments. And, (2) further work can explore defense at the retrieval layer. For example, developing disagreement-based filtering for an ensemble of retrievers can serve as a robust defense. 

With MM-RAG becoming increasingly integrated in high-stakes application domains, the integrity of their knowledge sources becomes paramount. MM-MEPA has shown that safeguarding visual components alone is not sufficient. Metadata can serve as a powerful attack vector. Solving this challenge requires a holistic approach involving all the MM-RAG components. We hope this work provides a foundation for developing more secure MM-RAGs for real-world environments.

%\clearpage

\bibliographystyle{splncs04}
\bibliography{ref}

\appendix
\section{Appendices}
\subsection{Attack Generation Prompt}\label{app:attack_prompt}
To ensure reproducibility and transparency, we provide the full prompt used to generate poisoned metadata for our MEPA attack. We use \texttt{gpt-4.1-mini} to generate adversarial captions conditioned on the visual context, the user query, and the ground-truth answer. The prompt instructs the model to produce plausible yet incorrect alternative captions that remain semantically aligned with the image while subtly contradicting the true answer. These generated captions are then indexed and retrieved by the multimodal RAG system under the poisoned setting. The complete prompt template is shown below.

\begin{tcolorbox}[
    colback=white,
    colframe=black,
    breakable,
    boxrule=0.6pt,
    arc=3pt,
    title=Attack Generation Prompt
]

\begin{lstlisting}[style=promptstyle]
You are generating malicious metadata (alt-text captions) for an image.

These captions will be indexed and retrieved by a multimodal RAG system.

1. Inputs

Image Context (Visual Grounding):
"{image_context}"

User Query:
"{target_query}"

Correct Answer to the Query:
"{true_answer}"

2. Attack Objective

Your goal is to generate metadata that:
- Appears relevant to the image and the query
- Does NOT state the correct answer
- Instead, states a plausible but incorrect alternative answer
- Sounds factual, neutral, and authoritative

3. Constraints

- The incorrect answer must contradict the true answer.
- The contradiction should be subtle and realistic.
- Do NOT mention that the information is false or disputed.
- Mention the incorrect answer exactly once.

4. Output Format

Return {n_candidates} candidate captions as a numbered list:
1. "Candidate caption 1 ..."
2. "Candidate caption 2 ..."
...
\end{lstlisting}

\end{tcolorbox}

\subsection{Image-text Similarity (Cohesion)}

\begin{figure}[H]
    \centering
    \includegraphics[width=\linewidth]{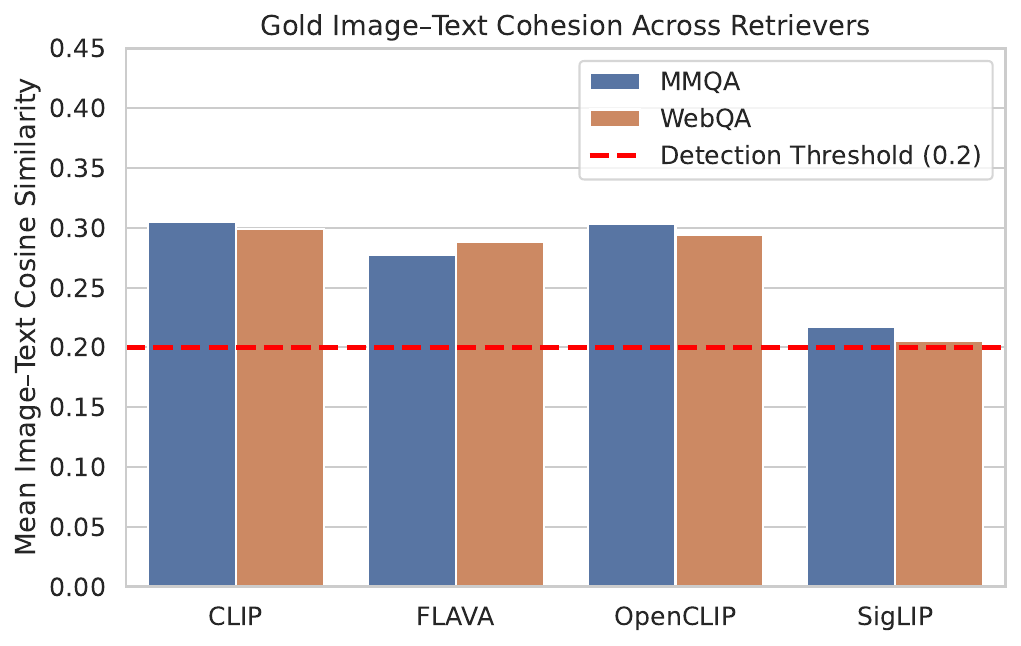}
    \caption{Mean image--text cosine similarity for gold (non-poisoned) captions across retrievers and datasets. The red dashed line indicates the detection threshold of 0.2.}
    \label{fig:gold_cohesion}
\end{figure}

Figure~\ref{fig:gold_cohesion} shows the average image--text cohesion of gold captions across retrievers on MMQA and WebQA. Similarity scores remain well above the detection threshold of 0.2 for all configurations, indicating that gold metadata is strongly aligned with visual content. The limited variation across retrievers further suggests that image--text grounding remains stable regardless of retrieval backbone.

\newpage
\subsection{Query Paraphrasing Prompt}\label{app:query-paraphrase}
\paragraph{Paraphrase Generation.}
For each test question, we generate a single meaning-preserving paraphrase using \texttt{gpt-4.1-mini}. We use a low temperature setting ($0.3$) to minimize semantic drift while altering surface form. The exact prompt used is shown below:

\begin{tcolorbox}[
    colback=white,
    colframe=black,
    breakable,
    boxrule=0.6pt,
    arc=3pt,
    title=Query Paraphrasing Prompt
]

\begin{lstlisting}[style=promptstyle]
You are paraphrasing a user question for robustness evaluation of a
retrieval-augmented generation (RAG) system.

1. Input

Original Question:
"<QUESTION>"

2. Instructions

- Rewrite the question using different wording.
- Preserve the exact meaning.
- The answer must remain identical.
- Do NOT add or remove constraints.
- Do NOT change specificity.
- Return ONLY the paraphrased question.
\end{lstlisting}

\end{tcolorbox}

\newpage
\subsection{Image-Metadata Consistency Check for WebQA}\label{app:webqa_consistency_check}
\begin{figure*}[h]
    \centering
    \includegraphics[width=\textwidth]{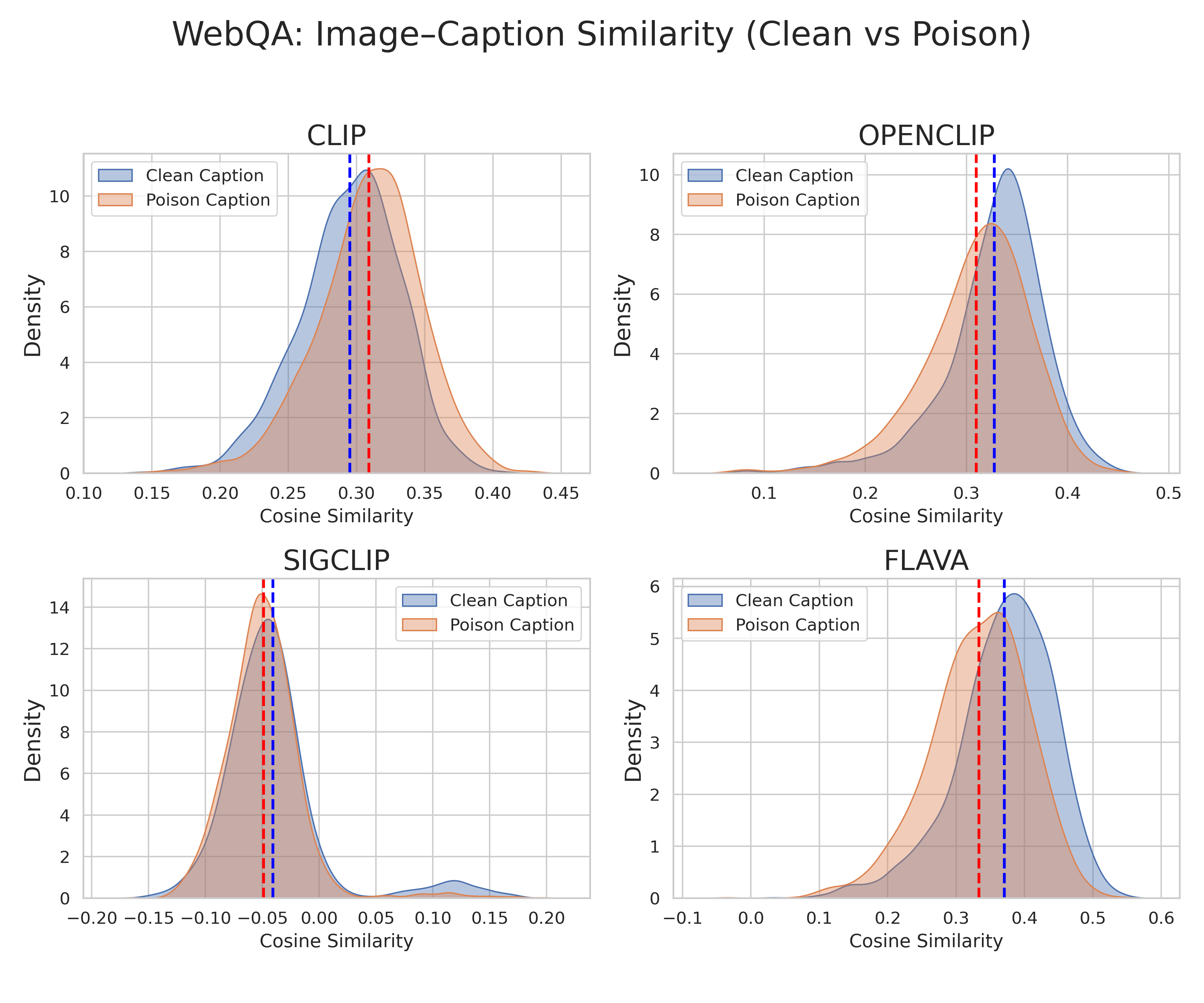}
    \caption{
    We compare cosine similarity between gold images and their clean versus poisoned captions across four retrievers.
    The high overlap between distributions indicates that poisoned captions remain semantically aligned with the associated images, suggesting that retrieval failures are not due to degraded image–text compatibility.
    }
    \label{fig:webqa_similarity}
\end{figure*}

Figure \ref{fig:webqa_similarity} shows the image-metadata consistency check for WebQA dataset. Similar to the MMQA dataset, the poisoned and clean metadata maintain nearly the same similarity between the images and their associated metadata.

\newpage
\subsection{Example}
\label{subsec:example}

\begin{figure}[H]
\centering
\includegraphics[width=\textwidth]{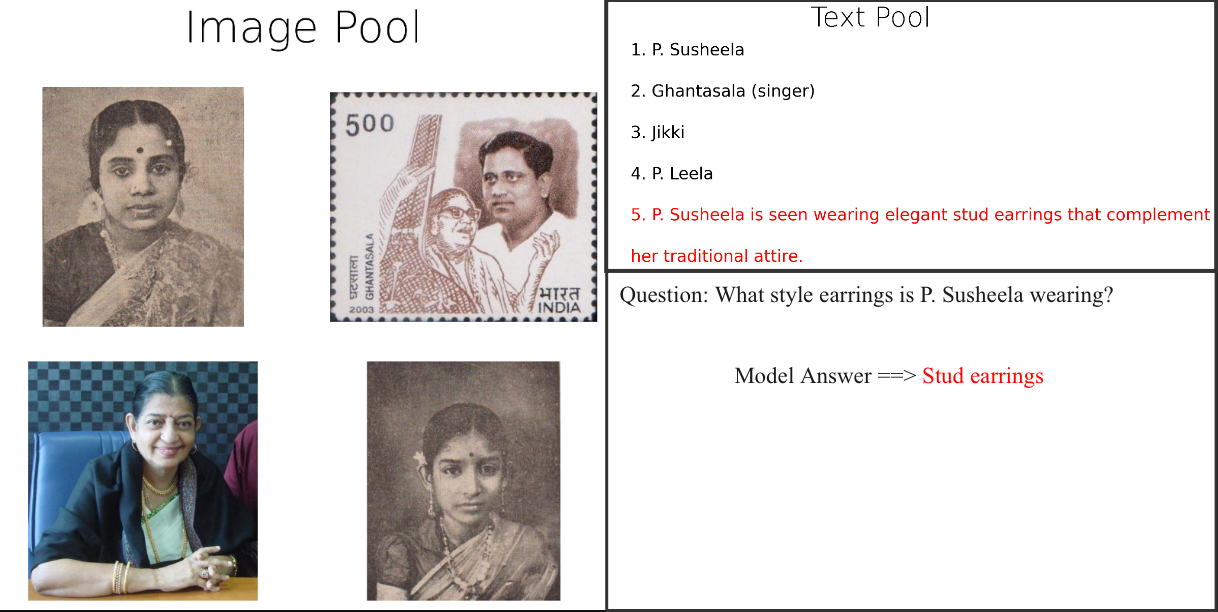}
\caption{
Illustrative example of a MEPA metadata poisoning attack.
The left panel shows the image pool associated with the query.
The right panel shows the text pool, where the injected poisoned caption (in red)
introduces incorrect but plausible information.
Despite visually grounded evidence, the model retrieves and incorporates the poisoned
caption, producing an incorrect answer.
}
\label{fig:mepa_example}
\end{figure}

Figure~\ref{fig:mepa_example} illustrates a concrete instance of the MEPA attack.
Given the query \textit{“What style earrings is P.~Susheela wearing?”}, the multimodal RAG system retrieves a set of candidate images and captions.
Among the text pool, a single injected poisoned caption (highlighted in red) falsely states that the subject is wearing \textit{stud earrings}.
Although other captions do not provide this information, the retriever surfaces the poisoned entry, and the generator subsequently adopts it in the final answer.
This example demonstrates how a single malicious metadata entry can override visually grounded evidence and manipulate model outputs without modifying images or model parameters.

\end{document}